\documentclass[10pt,twocolumn,preprintnumbers,superscriptaddress,nofootinbib,aps,prl]{revtex4-1}

\usepackage{graphicx}
\usepackage{amsmath}
\usepackage[utf8]{inputenc}
\usepackage[colorlinks,pdfusetitle]{hyperref}

\newcommand{\beq}{\begin{equation}}
\newcommand{\eeq}{\end{equation}}

\newcommand{\appropto}{\mathrel{\vcenter{
  \offinterlineskip\halign{\hfil$##$\cr
    \propto\cr\noalign{\kern2pt}\sim\cr\noalign{\kern-2pt}}}}}
\newcommand{\e}[1]{\times10^{#1}}
\newcommand{\orcid}[1]{\href{https://orcid.org/#1}{#1}}

\begin{document}

\pagestyle{plain}

\title{Ultra Light Boson Dark Matter and Event Horizon Telescope Observations of M87*}

\author{Hooman Davoudiasl}
\email{hooman@bnl.gov}
\thanks{\orcid{0000-0003-3484-911X}}

\author{Peter B. Denton}
\email{pdenton@bnl.gov}
\thanks{\orcid{0000-0002-5209-872X}}

\affiliation{Department of Physics, Brookhaven National Laboratory,
Upton, NY 11973, USA}


\begin{abstract}

The initial data from the Event Horizon Telescope (EHT) 
on M87$^*$, the supermassive black hole at 
the center of the M87 galaxy, provide direct observational information on its mass, spin, and accretion disk properties.  A combination of the EHT data and other constraints provide evidence 
that M87$^*$ has a mass $\sim 6.5 \times 10^9\,M_\odot$ and dimensionless spin parameter $|a^*|\gtrsim 0.5$.  These determinations disfavor ultra light bosons of mass $\mu_b\sim 10^{-21}$ eV, within the range considered for fuzzy dark matter, invoked to explain dark matter distribution on $\sim$ kpc scales.  Future observations of M87$^*$ could be expected to strengthen our conclusions.

\end{abstract}
\maketitle

\section{Introduction}
Black holes (BHs) are at the same time simple and mysterious.  They are characterized by only 
a few parameters - mass, spin, and charge - and are considered purely gravitational objects.  
Yet their essential character is quite enigmatic: they represent a one-way exit (up to quantum effects \cite{Hawking:1974sw})
from the causally connected Universe, and their internal properties  are masked by an event horizon 
that is the point of no return.  The most direct evidence for their existence has until very recently 
been provided by the observation of gravitational waves from binary mergers ascribed to black 
	holes \cite{Abbott:2016blz}.  This situation changed upon the release of a first ever image of the M87$^*$ supermassive black hole (SMBH) at the center of the Messier 87 (M87) galaxy, by the Event Horizon Telescope (EHT) \cite{Akiyama:2019cqa}.  In some sense, this is the most direct  evidence for BHs, as it 
manifests their defining characteristic: 
{\it a region of space from which no matter and light can escape}.

The EHT imaging of M87$^*$ through a worldwide network of radio telescopes is a historic scientific 
accomplishment.  Future observations of this and other SMBHs will usher in a new age of radio astronomy where direct data on their 
event horizons and associated accretion dynamics become available and 
will get increasingly more precise.  There are numerous astronomical questions that could be 
addressed with such observations and we can expect new and interesting questions to arise as well.  
However, it is also interesting to inquire whether the impressive new EHT data on M87$^*$ could also 
be used to shed light on fundamental questions of particle physics and cosmology.  

In this letter, we use the results of the EHT collaboration on the 
parameters of M87$^*$ in the context of particle physics, and in particular 
ultra light bosons.  These states could potentially provide motivated candidates for dark matter (DM), 
one of the most important open fundamental questions of physics.  Dark matter constitutes the dominant form 
of matter in the Universe, making up $\sim 25\%$ of its energy density \cite{Tanabashi:2018oca}, 
with at best feeble couplings  to the visible world.  It is generally assumed that  DM has non-gravitational interactions that led to its production in the early Universe.  These interactions could then result in its detection in a variety of laboratory experiments.  Nonetheless, DM  has only been observed through its gravitational effects in astrophysics and cosmology.  Therefore, 
purely gravitational probes of DM provide the most model-independent approach to constraining 
its properties.  

It turns out that BHs, themselves purely gravitational, can 
provide a unique probe of ultra light DM states through the mechanism of superradiance \cite{Penrose:1969pc,1971JETPL..14..180Z,Misner:1972kx,Press:1972zz,Press:1973zz,
Starobinsky:1973aij,Zouros:1979iw,Detweiler:1980uk,Brito:2015oca,Gates:2018hub}.  That is, 
roughly speaking, a spinning BH will lose its angular momentum very efficiently if a 
boson with a particular mass exists in the spectrum of physical states.  This is only a condition on the 
mass of the boson and does not depend on whether the boson has any non-gravitational interactions.  In fact, the boson does not even need to have any ambient number density, since quantum fluctuations suffice to populate a boson cloud around the BH by depleting its spin.

The superradiance mechanism
can then provide an interesting 
probe of DM states that would be otherwise practically inaccessible to experiments.  These states 
include ultra light axions \cite{Arvanitaki:2009fg,Arvanitaki:2014wva,Arvanitaki:2016qwi} and vector bosons \cite{Baryakhtar:2017ngi} that can appear in various high energy frameworks, such as  string theory.  For extremely small masses $\mu_b\sim10^{-(21-22)}~\text{eV}$ such states can also address certain observational 
features of the DM distribution on scales of $\sim$~kpc; this class of bosons is often referred 
to as fuzzy DM \cite{Hu:2000ke,Marsh:2015xka,Hui:2016ltb} (for a fuzzy DM model based on infrared dynamics, see Ref.~\cite{Davoudiasl:2017jke}).
We will show that the results of the EHT 
collaboration on the spin of the M87$^*$ SMBH \cite{Akiyama:2019fyp} 
can probe and constrain this interesting regime of ultralight DM masses.

\section{Superradiance Overview}
Black hole superradiance leads to the growth of the boson population once its energy $\omega_b$ satisfies the condition (see, for example, Refs.~\cite{Arvanitaki:2009fg,Arvanitaki:2014wva})
\beq
\frac{\omega_b}{m} < \Omega_H\,,
\label{SRcond}
\eeq
where $m$ is the magnetic quantum number of the boson, associated with its angular momentum.  Here, $\Omega_H$ is the angular velocity of the BH event horizon related to the dimensionless spin parameter $a^*\equiv J_{\rm BH}/(G_NM_{\rm BH}^2)\in[0,1)$ by 
\beq
\Omega_H = \frac{1}{2 r_g}\frac{a^*}{1 + \sqrt{1 - a^{*2}}}\,,
\label{OmegaH}
\eeq
where $r_g \equiv G_N M_{\rm BH}$, $G_N$ is Newton's constant, 
and $M_{\rm BH}$ is the BH mass.  In the above expression $J_{BH}$ is the BH angular momentum.

In addition to the condition in eq.~\ref{SRcond}, there is another condition that must be met for superradiance to deplete the spin of a BH,
\begin{equation}
\Gamma_b\,\tau_{\rm BH}\ge\ln N_m\,,
\end{equation}
where $\tau_{\rm BH}$ is the characteristic timescale of the BH,
$N_m$ is the final occupation number of the cloud after the BH spins down by $\Delta a^*$,
\begin{equation}
N_m\simeq\frac{G_NM_{\rm BH}^2\Delta a^*}{m}\,,
\end{equation}
and $\Gamma_b$ is the growth rate of the field for $b\in\{S,V\}$ (scalar or vector).
The leading contribution for $\Gamma_b$ is different for scalars and vectors and, up to a factor of $\sim2$, we have
\begin{align}
\Gamma_S&=\frac1{24}a^*r_g^8\mu_S^9\,,\\
\Gamma_V&=4a^*r_g^6\mu_V^7\,.
\end{align}

For an observation of a BH mass and spin, an upper and lower limit on $\mu_b$ can be placed (that is, demanding that superradiance has not depleted the spin of the BH by $\Delta a^*$) by,
\begin{equation}
\mu_b>\Omega_H\,,
\label{eq:mu limit}
\end{equation}
or
\begin{align}
\mu_S&<\left(\frac{24\ln N_m}{a^*r_g^8\tau_{\rm BH}}\right)^{1/9}\,,\label{eq:muS limit}\\
\mu_V&<\left(\frac{\ln N_m}{4a^*r_g^6\tau_{\rm BH}}\right)^{1/7}\,,\label{eq:muV limit}
\end{align}
where we used the fact that for the dominant mode one has $m=1$ for both scalars and vectors.
That is, if the constraint in eq.~\ref{eq:mu limit} applies to a larger mass than the constraint in eq.~\ref{eq:muS limit} or \ref{eq:muV limit}, the mass range of ultra light bosons in between is ruled out.

\section{EHT Observations of M87\texorpdfstring{$^*$}{*}}
The EHT has provided the first direct measurement of the environment immediately around M87$^*$, leading to a mass estimate of $(6.5\pm0.7)\e9\,M_\odot$ \cite{Akiyama:2019fyp}.
This is fairly consistent with previous estimates that are in the $[3.5,7.2]\e9\,M_\odot$ range \cite{Gebhardt:2011yw,Walsh:2013uua,2016MNRAS.457..421O}.

The shortest timescale that could be relevant for a SMBH is the Salpeter time $\tau_{\rm Salpeter}\sim4.5\e7$ years \cite{Shankar:2007zg} which is the case for when material is accreting on to the object at the Eddington limit.
In Ref.~\cite{Baryakhtar:2017ngi} they conservatively take $\tau_{\rm BH}\sim\tau_{\rm Salpeter}/10$ to account for the possibility of super-Eddington accretion.
Observations of M87$^*$, however, show that $\dot M/\dot M_{\rm Edd}\sim2.0\e{-5}$ \cite{Akiyama:2019fyp} consistent with previous measurements \cite{Kuo:2014pqa} which suggests that the relevant timescale is much longer.
We conservatively take $\tau_{\rm BH}=10^9$ years as our fiducial value since the accretion time is much longer.
In addition, in the last billion years there was likely only one merger event which involved a much smaller galaxy and was unlikely to significantly affect the spin of M87$^*$ \cite{2015A&A...579L...3L}.
We also note that the dependence of the ultra light boson limits on $\tau_{\rm BH}$ is at most $\tau_{\rm BH}^{-1/7}$.

The final parameter that remains to be observationally constrained, and perhaps the most important in this context, is the spin.
The EHT checked if various spin configurations are consistent with the data.
They found that $a^*=0$ is inconsistent with the data, while spins $|a^*|\ge0.5$ up to $|a^*|=0.94$ (as high as their analysis goes) are consistent with the data, although there was no analysis made of any spins $0<|a^*|<0.5$ \cite{Akiyama:2019fyp}.
This leads to an approximate estimate of $|a^*|>0.5$ which relies strongly on the observed jet power to rule out the smaller spins.
The EHT collaboration takes a very conservative estimate of the jet power \cite{Akiyama:2019fyp}.
A separate detailed analysis was performed which finds $a^*=0.9\pm0.1$ \cite{Tamburini:2019vrf} which we take as our fiducial value and uncertainty.

\begin{figure*}
\centering
\includegraphics[width=\textwidth]{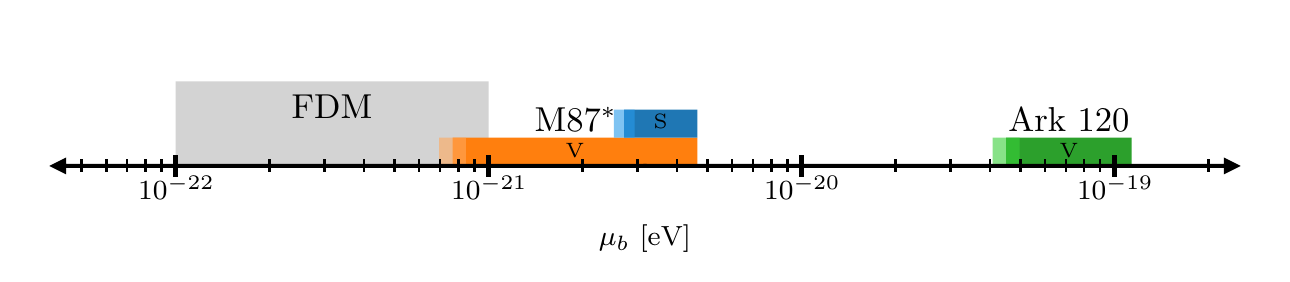}
\caption{Regions of parameter space constrained by observations of SMBHs.
The orange (blue) region is ruled out for vector (scalar) bosons by M87$^*$.
Each constraint is derived using the $1\,\sigma$ conservative values for the mass and spin, and the shaded band on the left of each region represents the size of the theoretical uncertainty.
The green region is the constraint on vector bosons from Ark 120 which cannot constrain scalars.
The region preferred by fuzzy DM is shown in gray.}
\label{fig:1d}
\end{figure*}

\section{Results}
Using eqs.~\ref{eq:mu limit}, \ref{eq:muS limit}, and \ref{eq:muV limit}, it is possible to constrain light bosons across a range of masses.
We assume that the largest value of $\Delta a^*$ is $1-a^*$ where $a^*$ is the spin today.
We report the $1\,\sigma$ results accounting for the uncertainties in the mass and spin as described in the previous section, as well as a factor of two in the uncertainty in the theoretical calculation of $\Gamma_b$.
Then we find that M87$^*$ rules out light bosons in the following ranges,
\begin{align}
2.9\e{-21}{\rm\ eV}&<\mu_S<4.6\e{-21}{\rm\ eV}\,,\\
8.5\e{-22}{\rm\ eV}&<\mu_V<4.6\e{-21}{\rm\ eV}\,,
\end{align}
as shown in fig.~\ref{fig:1d} which also includes the constraint from the lighter Ark 120 with $M_{\rm BH}=0.15\e9\,M_\odot$ and $a^*=0.64$ \cite{Baryakhtar:2017ngi,Reynolds:2013qqa,Walton:2012aw,Peterson:2004nu}.
For larger boson masses, there is fairly continuous coverage from $\mathcal O($few$)\e{-20}$ eV to $\mathcal O($few$)\e{-17}$ eV from SMBH observations with just a small gap at $\mathcal O($few$)\e{-19}$ eV.
Then there is large gap up to $\mathcal O($few$)\e{-14}$ eV at which point stellar mass BHs provide constraints up to $\sim10^{-11}$ eV.

\begin{figure}
\centering
\includegraphics[width=\columnwidth]{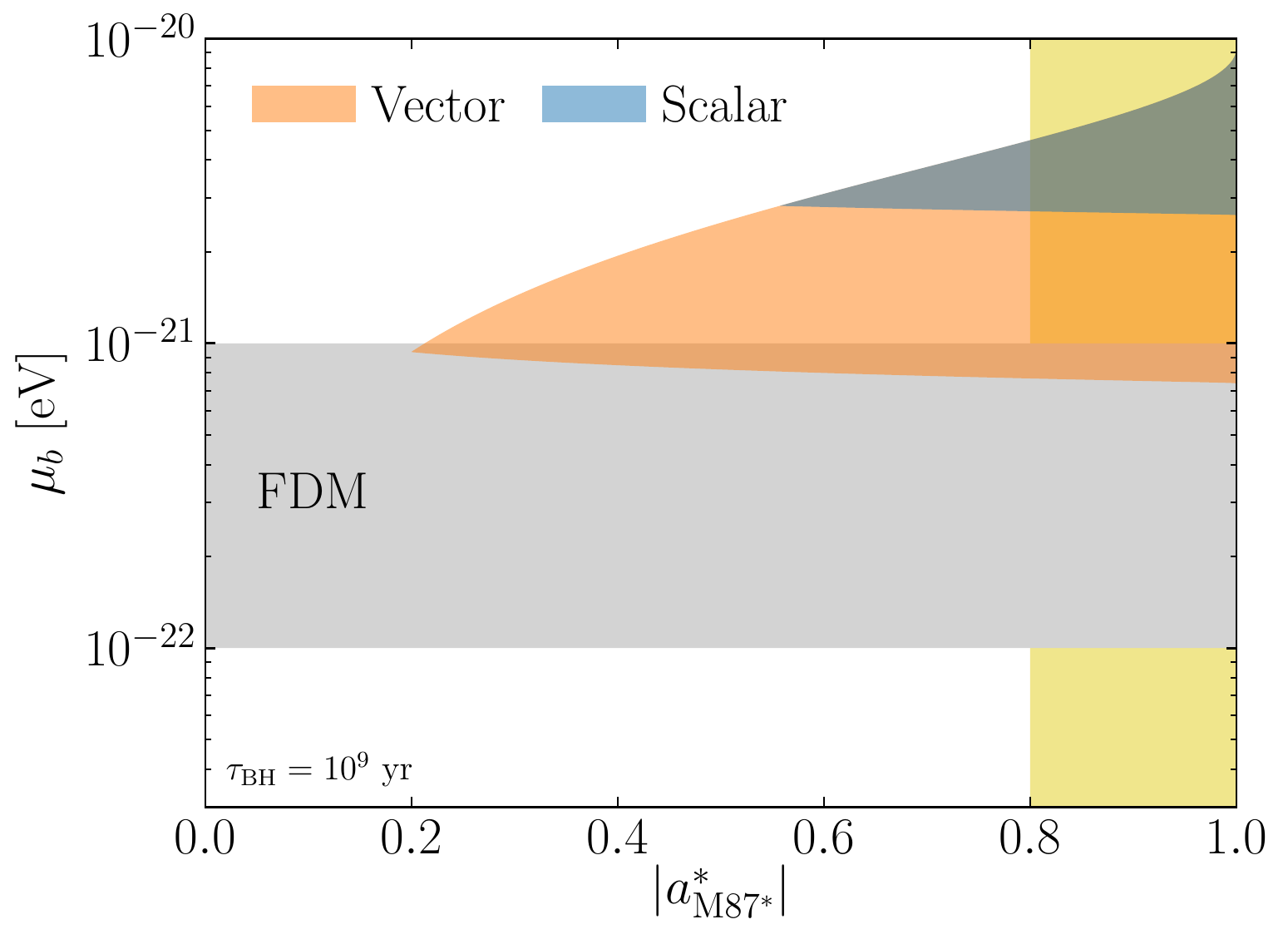}
\caption{The constraints on light bosons as a function of the spin of M87$^*$.
The region constrained for scalar bosons (blue) is also constrained for vector bosons (orange).
The characteristic fuzzy DM range is shown in gray, and the $1\,\sigma$ inference region of the spin is shown in khaki \cite{Tamburini:2019vrf}.}
\label{fig:a dependence}
\end{figure}

We also explored the effect of the spin measurement of M87$^*$ on the constraint, as shown in fig.~\ref{fig:a dependence}.
A constraint on vector bosons exists for any $|a^*|>0.2$ which overlaps with the fuzzy DM range.
A constraint on scalar bosons only exists for $|a^*|>0.55$, none of which probes the fuzzy DM range.

\section{Outlook}
With additional analyses and observations, the spin of M87$^*$ will become more precisely determined.
If the spin is on the larger end, the constraints on ultra light bosons will become stronger and for $|a^*|>0.55$ scalar bosons can also be constrained.

The largest SMBHs are more than an order of magnitude more massive than M87$^*$, but are significantly farther away making them difficult targets for the EHT or other probes that could provide good spin measurements \cite{Shemmer:2004ph}.
Still, this means that it is, in principle, possible to probe the entire fuzzy DM parameter spin using this technique, depending on the spins of the largest SMBHs.

Lyman-$\alpha$ forest measurements and observations of the heating of the core of star clusters provide separate constraints on fuzzy DM that disfavor most of the parameter space, leaving a possible opening around $\gtrsim10^{-21}$ eV \cite{Viel:2013apy,Irsic:2017yje,Marsh:2018zyw}.
We note that this region is now constrained by M87$^*$.

\section{Conclusions}
The Event Horizon Telescope has provided the first direct image of a BH.
We have shown that the information gained from this observation can be used to place constraints on particle physics, specifically ultra light bosons via the mechanism of superradiance.
Superradiance leads to a large extraction of energy from a rotating BH and any determination of a BHs spin could place a constraint on the presence of ultra light bosons.
The measurement of M87$^*$ provides constraints on both vector and scalar bosons, and in the vector case constrains some of the fuzzy dark matter parameter space.
Future observations of M87$^*$'s spin can pin down the exact constraint and, in principle, future spin measurements of SMBHs could possibly cover the entire fuzzy dark matter parameter space.

\begin{acknowledgments}
Work supported by the US Department of Energy under Grant Contract DE-SC0012704.
PBD acknowledges the hospitality of the Penn State physics department.
\end{acknowledgments}

\bibliography{m87-refs}

\end{document}